\documentclass[12pt]{article}
\pdfoutput=1
\usepackage{graphicx}
\usepackage{amssymb}
\usepackage{amsthm}
\usepackage{amsmath}
\usepackage[left=1.2in,right=1.2in,top=1.2in,bottom=1.2in]{geometry}
\usepackage{mathtools}
\usepackage{caption}
\usepackage{hyperref}
\usepackage[caption=false]{subfig}
\usepackage{xcolor}
\hypersetup{
    colorlinks,
    linkcolor={red!50!black},
    citecolor={blue!50!black},
    urlcolor={blue!80!black}
}

\newcommand{\R}{\mathbb{R}}
\newcommand{\Z}{\mathbb{Z}}

\newcommand{\rmd}{\mathrm{d}}

\begin{document}

\begin{center}
\vspace{24pt}
{ \large \bf Geodesic distances in Liouville quantum gravity}

\vspace{30pt}

{\sl J. Ambj\o rn}$\,^{a,b}$
and {\sl T.G. Budd}$\,^{a}$

\vspace{48pt}
{\footnotesize

$^a$~The Niels Bohr Institute, Copenhagen University\\
Blegdamsvej 17, DK-2100 Copenhagen \O , Denmark.\\
{ email: ambjorn@nbi.dk, budd@nbi.dk}\\

\vspace{10pt}
$^b$~Institute for Mathematics, Astrophysics and Particle Physics (IMAPP)\\ 
Radbaud University Nijmegen, Heyendaalseweg 135,
6525 AJ, Nijmegen, The Netherlands 
}
\vspace{36pt}
\end{center}


\begin{center}
{\bf Abstract}
\end{center}
\noindent
In order to study the quantum geometry of random surfaces in Liouville gravity, we propose a definition of geodesic distance associated to a Gaussian free field on a regular lattice. 
This geodesic distance is used to numerically determine the Hausdorff dimension associated to shortest cycles of 2d quantum gravity on the torus coupled to conformal matter fields, showing agreement with a conjectured formula by Y. Watabiki.
Finally, the numerical tools are put to test by quantitatively comparing the distribution of lengths of shortest cycles to the corresponding distribution in large random triangulations.
\vspace{12pt}
\noindent

\vspace{24pt}
\noindent
PACS: 04.60.Ds, 04.60.Kz, 04.06.Nc, 04.62.+v.\\
Keywords: Liouville gravity, geodesic distance, Gaussian free field, random surfaces.

\section{Introduction}\label{sec:intro}

Some 25 years ago the path integral of two-dimensional Euclidean quantum gravity coupled to conformal matter fields with central charge $c \leq 1$ was computed using a conformal bootstrap \cite{knizhnik_fractal_1988,david_conformal_1988,distler_conformal_1989}.
In this approach, known as \emph{Liouville quantum gravity}, the quantum geometry is described by a relatively simple quantum field theory, allowing various observables to be calculated analytically (see, e.g., \cite{seiberg_notes_1990,francesco_2d_1995,ginsparg_lectures_1993,nakayama_liouville_2004} for reviews).
Parallel to these developments in the continuum, much progress was made in studying two-dimensional quantum gravity using lattice models known as \emph{dynamical triangulations} or \emph{matrix models} (see, e.g., \cite{ambjorn_quantum_1997} for an overview).
These lattice theories rely on sampling large random triangulations, or similar combinatorial objects, and many quantities can be computed exactly for finite lattice spacing.
Quite remarkably, whenever an observable could be computed both in Liouville quantum gravity and in a lattice theory, agreement was found as the lattice spacing was taken to zero.
This strongly suggests that the continuum limit of these lattice models is in fact Liouville quantum gravity.
Proving this statement in one form or the other has become one of the main goals for both physicists and mathematicians working on two-dimensional gravity.

Several recent mathematical breakthroughs have brought us closer to this goal.
On the Liouville quantum gravity side, there has been significant progress in rigorously constructing the so-called quantum Liouville measure 
\cite{duplantier_liouville_2011}, to be described in the next section. 
On the lattice side, the continuum limit of random planar triangulations has been identified as a universal random metric space, known as the \emph{Brownian map} \cite{marckert_limit_2006,gall_uniqueness_2013,miermont_brownian_2013}.

Together, the formulations of two-dimensional Euclidean quantum gravity have a history of being mutually inspirational, in the sense that problems appearing on one side often turn out to be more easily solved on the other.
This interaction between Liouville quantum gravity and random triangulations will play an important role in the work reported here. 
Our main goal is to make sense of geodesic distances associated to the quantum geometry defined by the metric
\begin{equation}\label{eq:j1}
g_{ab}(x) = e^{\gamma \phi(x)} \hat{g}_{ab}(x;\tau),
\end{equation}
where $\phi$ is a quantum Liouville field and $\hat{g}_{ab}(x;\tau)$ is a fixed classical background metric. 
Equation (\ref{eq:j1}) has a clear interpretation in terms of Riemannian geometry as long as one puts an ultraviolet cut-off on the Liouville field.
However, it is still an open question how one should remove the cut-off in order to obtain a genuinely continuum random metric space. 

On the lattice side, there is no difficulty in defining geodesic distances, since there are natural graph distances associated to triangulations.
In the case of ''pure gravity'' these geodesic distances have been proven to have dimension of volume to the power $1/4$, implying a Hausdorff dimension $d_h=4$.
This shows that quantum effects are very important and can lead to fractal geometry.
Currently the necessary analytic tools are lacking in order to compute distances on the lattice when coupled to conformal matter fields.
However, there is a conjectured formula relating the Hausdorff dimension to the central charge $c$ of the matter fields based on a calculation in Liouville quantum gravity \cite{watabiki_analytic_1993},
\begin{equation}\label{eq:watdhofc}
d_h = 2 \frac{\sqrt{25-c}+\sqrt{49-c}}{\sqrt{25-c}+\sqrt{1-c}},
\end{equation}
which for $c=-2$ was shown to agree numerically with the scaling of distances in spanning-tree-decorated triangulations \cite{kawamoto_fractal_1992,ambjorn_quantum_1998}.

In this paper we will propose a definition for the geodesic distance associated to a quantum Liouville field, which can be studied numerically by putting the Liouville field on a regular lattice.
This allows us not only to test the proposed definition, but also to measure the associated Hausdorff dimension.
One advantage of this method of determining the Hausdorff dimension, as compared to using random triangulations, is that we can test formula (\ref{eq:watdhofc}) for a much larger range of central charges $c$, since, as we will see in the following, the central charge $c$, or rather the associated scaling exponent $0<\gamma<2$, is a freely adjustable parameter in the simulations.

\section{Liouville quantum gravity on the torus}

Liouville quantum gravity arises from gauge fixing the path integral over two-dimen\-sional geometries coupled to conformal matter fields.
Its partition function for a surface $S$ of genus $g$ can be formally written as \cite{david_conformal_1988,distler_conformal_1989}
\begin{equation}\label{eq:part1}
Z = \int_{\mathcal{M}_g} \rmd\tau\, Z(\tau),\quad Z(\tau) = \int \mathcal{D}_{\hat{g}_{\tau}} \phi\int \mathcal{D}_{\hat{g}_{\tau}} X\, J_{\hat{g}_{\tau}} \exp(- S_L[\phi,\hat{g}_{\tau}] - S_m[X,\hat{g}_{\tau}]),
\end{equation}
which includes an integration over a family of background metrics $\hat{g}_{\tau}$ parametrized by  coordinates $\tau$ on the genus-$g$ Moduli space $\mathcal{M}_g$, a functional integration over the Liouville field $\phi : S\to \R$, and another functional integration over a set of matter fields $X$.
Both $J_{\hat{g}_{\tau}}$, which comes from the Faddeev-Popov determinant associated
with the gauge fixing to conformal gauge,  
and the matter action $S_m[X,\hat{g}_{\tau}]$ 
depend non-trivially on the background metric $\hat{g}_{\tau}$, 
but are independent of the Liouville field.
Therefore, if one is only interested in the quantum 
geometry of the surface $S$, one may perform the integral over 
the matter fields and (\ref{eq:part1}) reduces to 
\begin{equation}\label{eq:part2}
Z(\tau) = \rho(\tau) \int \mathcal{D}_{\hat{g}_{\tau}}\phi \exp(- S_L[\phi,\hat{g}_{\tau}]),
\end{equation}
for some function $\rho(\tau)$ and $S_L[\phi,\hat{g}_{\tau}]$ is the Liouville action
\begin{equation}
S_L[\phi,\hat{g}] = \frac{1}{4\pi} \int \rmd^2x\sqrt{\hat{g}(x)}\left(\hat{g}^{ab}\partial_a\phi\partial_b\phi + Q \hat{R} \phi + 4\pi \lambda e^{\gamma \phi}\right),
\end{equation}
where $\hat{R}$ is the scalar curvature of the background metric and $\lambda$ is the cosmological constant.
According to the conformal bootstrap approach, the requirement that the partition function (\ref{eq:part1}) is independent of the family of background metrics $\hat{g}_{\tau}$ fixes the parameters $Q$ and $\gamma$ in terms of the central charge $c$ of the matter fields $X$,
\begin{equation}
Q = \frac{2}{\gamma} + \frac{\gamma}{2} = \sqrt{\frac{25-c}{6}}.
\end{equation} 

In the following we will restrict our attention to a surface of genus $g=1$, for which the partition function becomes particularly simple. 
One may choose the background metrics to be flat, i.e., $\hat{R}=0$, and parametrized by a complex modulus $\tau = \tau_1 + i \tau_2$ in the standard way (see also figure \ref{fig:dtloop}),
\begin{equation}
\hat{g}_{ab}(\tau) = \frac{1}{\tau_2} \begin{pmatrix} 1 & \tau_1 \\ \tau_1 & \tau_1^2+\tau_2^2 \end{pmatrix}.
\end{equation}
With this choice and using an inverse Laplace transform to switch to the fixed-volume partition function, one obtains
\begin{align}
Z &= \int_{\mathcal{M}_1} \!\!\rmd \tau \int_0^{\infty} \!\!\rmd V e^{-\lambda V} Z(\tau,V), \\ 
Z(\tau,V) &= \rho(\tau) \int \mathcal{D}_{\hat{g}_{\tau}}\phi\, \delta\left(V - \int \rmd^2x\, e^{\gamma \phi}\right) \exp\left[-\frac{1}{4\pi} \int \rmd^2x\, \hat{g}^{ab}(\tau)\partial_a\phi\partial_b\phi\right].
\end{align}
One can take care of the delta function by integrating over the zero mode $\phi_0$ of $\phi(x) = \phi_0 + h(x)$,
\begin{equation}\label{eq:partgff}
Z(\tau,V) = \rho(\tau) \int \mathcal{D}_{\hat{g}(\tau)}h\,\exp\left[-\frac{1}{4\pi} \int \rmd^2x\, \hat{g}^{ab}(\tau)\partial_ah\partial_bh\right], \quad \int \rmd^2x\, h(x) = 0,
\end{equation}
which is simply the partition function of a Gaussian free field on a flat torus.

This suggests that one may construct a random metric $g_{ab} = e^{\gamma\phi}\hat{g}_{ab}(\tau)$ on the torus of desired volume $V$ in the following way: first one samples $\tau$ from the measure $Z(\tau)\rmd \tau$ on moduli space\footnote{A closed form expression is known for $Z(\tau)$ in (\ref{eq:part2}) in terms of the Dedekind $\eta$-function, $Z(\tau) = \tau_2^{-2} \left(\sqrt{\tau}|\eta(\tau)|^{2}\right)^{1-c}$. See e.g., \cite{gupta_random_1990}.}, then one samples a Gaussian free field $h$ on $\hat{g}_{ab}$, and finally one obtains the Liouville field $\phi$ by shifting the constant mode such that $\int \rmd^2x\, e^{\gamma \phi}=V$, i.e.
\begin{equation}\label{eq:phinorm}
\phi(x) = h(x) + \frac{1}{\gamma} \log\left(\frac{V}{\int\rmd^2y\,e^{\gamma h(y)}}\right).
\end{equation} 
Of course, this procedure is ill-defined without proper regularization of the Gaussian free field $h(x)$. 

It is still an open question how to assign a metric interpretation to $g_{ab} = e^{\gamma\phi}\hat{g}_{ab}(\tau)$, but a rigorous definition of the measure $\sqrt{g}\rmd^2x$ has been put forward in \cite{duplantier_liouville_2011}.
Let us briefly summarize the construction in the case of the torus $S$ equipped with the standard Lebesgue measure.
The values of a Gaussian free field $h(x)$ have infinite variance, but the random variable obtained by integrating $h$ w.r.t. a smooth function $f:S\to\R$, i.e., $\int_S \rmd^2x\,f(x)h(x)$, is almost surely finite.
Let $\{f_i:S\to\R\}_{i=1}^{\infty}$ be a sequence of smooth function forming an orthonormal basis of the Hilbert space closure of the space of smooth functions on $S$ with respect to the standard inner product.
Then one can define the projection $h_n(x)$ of $h(x)$ onto the subspace spanned by the first $n$ functions $\{f_i\}_{i=1}^{n}$.
Since $h_n(x)$ is a Gaussian random variable with zero mean, one finds
\begin{equation}
\langle e^{\gamma h_n(x)}\rangle = \exp\left(\gamma^2 \langle h^2_n(x)\rangle/2\right). 
\end{equation}
The variance $\langle h^2_n(x)\rangle$ grows without bound as $n\to\infty$, and therefore one needs to cancel it in order to have a chance of obtaining a continuum measure.
In fact, it is shown in \cite{duplantier_liouville_2011} that the measures\footnote{We disregard the contribution from the conformal radius that appears in the definition in \cite{duplantier_liouville_2011}, because in the case of the flat torus it only affects the overall normalization of the measure. } 
\begin{equation}
\rmd\mu_n := \exp\left( \gamma h_n(x) - \gamma^2 \langle h^2_n(x)\rangle/2 \right)\rmd^2x
\end{equation}
converge almost surely (weakly) to a well-defined measure $\rmd\mu_\gamma$, the \emph{quantum Liouville measure}, on $S$ as $n\to\infty$, which is independent of the chosen basis $\{f_i\}_{i=1}^{\infty}$.
Likewise, one has the \emph{normalized quantum Liouville measure}
\begin{equation}
\rmd \mu_{\gamma,V} := \lim_{n\to\infty} V \frac{\rmd \mu_n}{\int_S \rmd\mu_n}.
\end{equation}

Since for each $n$ the function $h_n$ is smooth, it is tempting to consider the corresponding smooth geometries defined by the Riemmannian metrics 
\begin{equation}\label{eq:naivemetric}
\exp\left( \gamma h_n(x) - \gamma^2 \langle h^2_n(x)\rangle/2 \right) \hat{g}_{ab}(\tau).
\end{equation}
However, it is not expected that the geodesic distances computed with respect to this metric converge as $n\to\infty$.
Intuitively this can be understood as follows. 
The measures $\rmd\mu_n$ are normalized in such a way that for a given region $A$, for $n$ large enough the measure of $A$ becomes approximately independent of $n$.
However, if one looks at how the measure is distributed within $A$, one will notice that it gets redistributed on small scales and the measure picks up more and more fine-grained structure.
This means that a shortest path with respect to the metric (\ref{eq:naivemetric}) which traverses $A$ will have more and more valleys, i.e., subsets of $A$ where the measure density is smaller than average, to choose from as $n$ increases.
Therefore one expects the geodesic distance to keep decreasing and to approach zero as $n\to\infty$.\footnote{In a rigorous setting the degeneracy of the intrinsic metric associated to a natural Dirichlet form was proven in \cite{garban_heat_2013}.}
At least a renormalization of the metric is necessary, e.g., by changing the factor $\gamma^2/2$ in the exponential in (\ref{eq:naivemetric}), to obtain finite distances in the $n\to\infty$ limit.
More importantly, there is no reason to believe that a possible limit of the geodesic distances with respect to (\ref{eq:naivemetric}) is independent of the chosen basis $\{f_i\}_{i=1}^{\infty}$.

A natural choice of basis $\{f_i\}_{i=1}^{\infty}$ is the set of eigenmodes of the Laplacian $\hat{\Delta}$ of the background metric ordered by (increasing absolute) eigenvalue, corresponding to a uniform momentum cut-off on the Liouville field. 
However, such a cut-off goes against the geometric spirit of 2d gravity, since it introduces a dependence on the background metric $\hat{g}$. 
Preferably one would introduce a so-called covariant cut-off that only depends on the physical metric $g_{ab}$, but a cut-off is needed exactly to define the latter, so it seems one is running in circles.
Luckily, one aspect of the physical metric is unambiguously defined, as we saw above, namely the measure $\rmd\mu_\gamma$. 
Instead of fiddling with the basis $\{f_i\}_{i=1}^{\infty}$ and the fields $h_n$, it is convenient to use the measure $\rmd\mu_\gamma$ as a starting point and to construct a metric by applying a ``filter'' to this measure\footnote{We use the word ``filter'' in analogy to the terminology in signal processing, where a filter generally refers to a process that removes an unwanted component from a signal. In fact, the filter we will describe shortly can be understood as a non-linear generalization of the standard filters used in image processing to ``blur'' images.}.

\begin{figure}[t]
\begin{center}
\includegraphics[width=.9\linewidth]{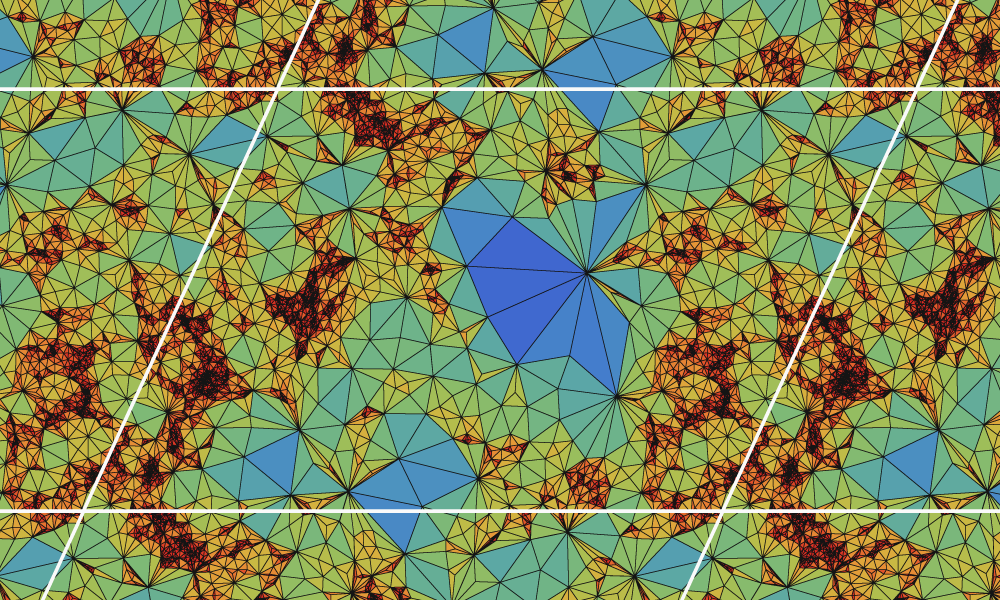}
\end{center}
\caption{Discrete harmonic embedding of a triangulation of 30000 triangles. The triangulation was chosen uniformly at random among all triangulations with a marked spanning tree (which is not shown). Such a random spanning-tree-decorated triangulation has the interpretation as gravity coupled to matter fields with central charge $c=-2$ (see also the discussion in section \ref{sec:cycles}). }%
\label{fig:harm}
\end{figure}

To get an idea what such a filtered measure could look like, let us examine how it is implicitly realized in the setting of random triangulations.
To a combinatorial triangulation one can assign a piece-wise flat Riemannian geometry by taking all triangles to be identical and equilateral.
According to the Riemannian uniformization theorem, for a triangulation of the torus the resulting geometry can be uniquely conformally mapped to a flat torus.
Various techniques are available to approximate this conformal map using discrete methods, e.g., via \emph{circle packings} or \emph{discrete harmonic embedding}, also known as Tutte embedding.
In the case of the torus the discrete harmonic embedding is computationally quite convenient.
It amounts to positioning the vertices of the triangulation in a flat torus in such a way that each vertex is located at the center of mass of its neighbors, see figure \ref{fig:harm} for an example.  
For more details on this embedding and determination of the corresponding modulus $\tau$, we refer the reader to \cite{ambjorn_roaming_2011}.

The discrete conformal map naturally associates to a triangulation a measure on a flat torus, namely the push-forward of the measure corresponding to the piece-wise flat Riemannian geometry of the triangulation.
For a finite triangulation this measure possesses a cut-off, in the sense that it is uniform on sufficiently small scales, namely within the individual triangles. 
However, the length scale of the cut-off is not fixed with respect to the background metric, but is set by the sizes of the triangles. 
Since the measure of each triangle is identical, say equal to $\delta$, one can say that within each disk-shaped region of measure of the order $\delta$ the measure is approximately evenly distributed.

Distances within random triangulations are well-defined and are known to converge under suitable rescaling as the cut-off is removed, i.e., as the number of triangles is taken to infinity.
This fact, together with the assumption that the random measure defined by the random triangulation converges to a quantum Liouville measure, suggests the implementation of the following filter for the quantum Liouville measure: 
Given a small $\delta > 0$, subdivide the surface $S$ into approximately disk-shaped regions $\{A\}$ having measure $\delta$ each; define the filtered measure $\rmd\mu_\delta$ to be uniform within each region $A$ such that $\mu_\gamma(A) = \mu_{\delta}(A)$.

There are various ways to implement such a filter, some of which are studied numerically in the next section, but in general the measure $\rmd\mu_\delta$ will correspond to a finite positive density (compared to the Lebesgue measure).
Therefore one can write
\begin{equation}
\rmd\mu_{\delta} = \rho_{\delta} \,\rmd^2 x,
\end{equation}
for some positive function $\rho_{\delta}(x)$.
Interpreting $\rho_{\delta}(x)$ as the density of a Riemannian metric, one can define a geodesic distance in the usual way,
\begin{equation}\label{eq:dist}
d_{\delta}(x,y) := \inf_{\Gamma:[0,1]\to S} \int_0^1\rmd t \sqrt{\rho_{\delta}} \|\Gamma'(t)\|_{\hat{g}_\tau},
\end{equation}
where the infimum is over all rectifiable curves $\Gamma$ from $x$ to $y$, and $\|\cdot\|_{\hat{g}_\tau}$ is the norm with respect to the background metric.
It is natural to conjecture that under reasonable assumptions there exists a distance function $d : S\times S \to \R$ and a positive real number $d_h$ such that pointwise\footnote{Similar conjectures have been made in \cite{sheffield_conformal_2010}, section 1.2, and \cite{miller_quantum_2013}, section 3.3}
\begin{equation}\label{eq:distconv}
\lim_{\delta\to 0} \delta^{\frac{1}{d_h} - \frac{1}{2}} d_{\delta}(x,y) = d(x,y).
\end{equation}
The exponent $d_h$ is expected to be the Hausdorff dimension of the limiting metric space, although this does not directly follow from the convergence of (\ref{eq:distconv}) without additional assumptions on, say, the conformal properties of the metric space.\footnote{The Hausdorff dimension is usually defined in terms of the growth of the volume of a geodesic ball as function of its radius.} 
As mentioned in the introduction, a relation between the Hausdorff dimension and the central charge $c$ was conjectured in \cite{watabiki_analytic_1993}, which in terms of $\gamma$ can be written as
\begin{equation}\label{eq:watab}
d_h(\gamma) = 1 + \frac{\gamma^2}{4} + \sqrt{\left(1+\frac{\gamma^2}{4}\right)^2+\gamma^2}.
\end{equation}

We will not attempt to prove (\ref{eq:distconv}) or (\ref{eq:watab}) in this paper.
Instead, we will collect some numerical evidence in support of the conjecture by putting the Liouville field on a lattice.

\section{Gaussian free fields on the lattice}\label{sec:dgff}

The partition function (\ref{eq:partgff}) is easily discretized on a regular lattice, especially if one sets $\tau= i$, which we will do in the following.
For $w$ a positive integer, consider the discrete Gaussian free field $h : \Z/w\times\Z/w \to \R$ on the periodic square $w\times w$ lattice with partition function
\begin{equation}\label{eq:discretepart}
\int \rmd^{w^2}h \exp\left[ - \frac{1}{4\pi} \sum_{|x-x'|=1} (h(x)-h(x'))^2\right] \delta\left(\sum_x h(x)\right),
\end{equation} 
where the sum is over all unordered pairs of neighbouring lattice sites.
The corresponding discrete quantum Liouville measure $\mu_{\gamma,w}$ is then given per lattice site by
\begin{equation}\label{eq:discrmeas}
\mu_{\gamma,w}(x) := w^{-2-\gamma^2/2} e^{\gamma h(x)}
\end{equation}
and its normalized version is
\begin{equation}\label{eq:discrvolnorm}
\mu_{\gamma,V,w}(x) := V \frac{\mu_{\gamma,w}(x)}{\sum_y\mu_{\gamma,w}(y)}.
\end{equation}
Since the Fourier modes of $h$ are independent Gaussian variables, one can efficiently generate random fields according to the partition function (\ref{eq:discretepart}) by using a discrete Fourier transform.
For more details, see for instance section 4 of \cite{sheffield_gaussian_2003}. 
Unless stated otherwise, in the following the unnormalized measure (\ref{eq:discrmeas}) will be used for simplicity, but can substituted by the normalized measure if desired.

\begin{figure}[t]
\begin{center}
\includegraphics[width=.3\linewidth]{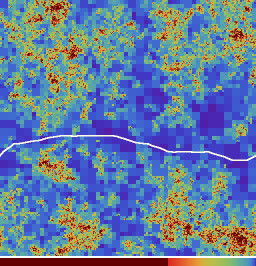}
\hspace{.01\linewidth}
\includegraphics[width=.3\linewidth]{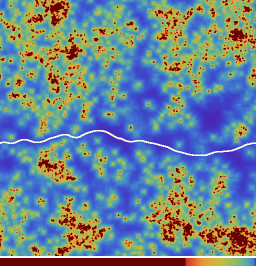}
\hspace{.01\linewidth}
\includegraphics[width=.3\linewidth]{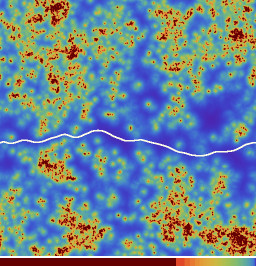}

(a)\hspace{.29\linewidth}(b)\hspace{.29\linewidth}(c)
\end{center}
\caption{Three different regularization methods. Each of the figures represents a regularized measure with $\gamma=1$, $\delta = 1/16000$, and $w=512$. The regularization methods are: (a) box subdivision, (b) box averaging, (c) disk averaging. The white curves represent the shortest cycles that will be discussed in section \ref{sec:cycles}.}%
\label{fig:reg}
\end{figure}

We would like to average the Liouville measure $\mu_{\gamma,w}(x)$ over regions of measure approximately $\delta$.
One particularly simple way to do this is by so-called \emph{box subdivision}, which was introduced in similar form in \cite{duplantier_liouville_2011}.
For this method to work one has to take the lattice $w$ to be a power of two.
Starting with a single square covering the whole lattice, i.e., having corners covering $(0,0)$ and $(w-1,w-1)$, one recursively subdivides each square according to the following criterion.
A square $s$ of edge length $2^k$ is subdivided into four smaller squares of edge length $2^{k-1}$ whenever $k\geq 1$ and the measure $\mu_{\gamma,w}(s) = \sum_{x\in s}\mu_{\gamma,w}(x)$ of $s$ exceeds $\delta$.
Otherwise, one sets $\mu_{\delta,\gamma,w}(x)$ for each lattice site $x$ in $s$ equal to the average value $2^{-2k}\mu_{\gamma,w}(s)$.
By construction the resulting discrete measure $\mu_{\delta,\gamma,w}$ has the same total measure as $\mu_{\gamma,w}$, but is coarser in regions where $\mu_{\gamma,w}(x)$ is relatively small.
An example of a random measure $\mu_{\delta,\gamma,w}$ with $\gamma = 1$, $\delta = 1/16000$, and $w=512$, is shown in figure \ref{fig:reg}a.
Although this method is computationally quite convenient, it has some downsides when one wants to consider the measure as an approximation to that of a smooth Riemannian metric. 
Indeed, the regularized measure is discontinuous and has significant jumps, especially for large $\gamma$, when passing between neighbouring squares.
Moreover, the resulting measure is genuinely anisotropic, even as the lattice size $w\to\infty$ for fixed $\delta$.

An alternative regularization, which we will refer to as \emph{disk averaging}, circumvents these issues in the following way.
For each site $x$ the radius $\epsilon_{x}$ of the disk centered at $x$ is determined that has measure (approximately) equal to $\delta$.
The regularized measure of site $x$ is then taken to be
\begin{equation}
\mu_{\delta,\gamma,w}(x) := \frac{\delta}{\pi \epsilon_{x}^2}.
\end{equation}
When $w$ is sufficiently large and $\delta$ sufficiently small, this regularized random measure will be (locally) isotropic, which is a necessary condition when one wants to trust outcomes of observables quantitatively.
The result of disk averaging is shown in \ref{fig:reg}c.

Unfortunately, the computation of the radii $\epsilon_{x}$ is quite time-consuming and does not allow us currently to go to much larger lattices than $w=512$.
As a compromise, for the measurements in section \ref{sec:cycles} a third method is used, referred to as \emph{box averaging}, where instead of determining a disk centered at a lattice $x$ one determines a square having measure $\delta$ centered at $x$.  
The box-averaged measure is not quite isotropic, but, as can be seen in figure \ref{fig:reg}b, it looks very similar to the disk-averaged measure.

\section{Discrete geodesic distance}\label{sec:geod}

Given a discrete measure $\mu_{\delta}(x)$, the simplest discretization of the geodesic distance (\ref{eq:dist}) is a weighted graph distance on the regular grid,
\begin{equation}\label{eq:discrdist}
d_{\delta}(x,y) := \inf_{t\to x_t} \sum_{t=1}^n \sqrt{\mu_{\delta}(x_t)},
\end{equation}
where the infimum is now over all discrete paths $(x_0=x,x_1,\ldots,x_n = y)$ on $(\Z/w)^2$ of arbitrary length $n$ such that $|x_{t+1}-x_{t}|=1$.
The problem with this definition is that the distance does not converge to the Riemannian distance when one would take $\mu_{\delta}$ successively finer approximations to the measure of a Riemannian metric.
In particular, if one takes $\gamma\to 0$ the random Liouville measure $\mu_{\gamma,\delta}$ becomes uniform and (\ref{eq:discrdist}) becomes proportional to the length of the shortest discrete path in the lattice.
This length corresponds to the  Manhattan distance between $x$ and $y$ which is quite different from the Euclidean distance.

As for the choice of averaging method, the anisotropy of the distance functions is not expected to affect the scaling properties, but to compare distances quantitatively to other approaches an asymptotically isotropic distance would be better.
Luckily, (\ref{eq:discrdist}) can be improved by solving a \emph{discrete eikonal equation} without significantly changing the complexity of its computation.
Details can be found in \cite{sethian_fast_1996,cohen_global_1997} (see chapter 2 of \cite{peyre_geodesic_2010} for a recent overview).
For a Riemannian metric $\rho(x)\rmd x^2$ the geodesic distance $d_{y}(x):=d(x,y)$ of $x$ to $y$ is a (weak) solution to the \emph{eikonal equation} 
\begin{equation}
\| \nabla d_y(x) \| = \frac{1}{\sqrt{\rho(x)}}\quad\text{ for }y\neq x \quad\text{and }\quad d_y(y)=0. 
\end{equation}
Discretization leads to the discrete Eikonal equation
\begin{equation}\label{eq:discreik}
d_y(x) = \min_{i\in\Z/4} v_{d_y}(x,x_i,x_{i+1})\quad\text{ for }y\neq x \quad\text{and }\quad d_y(y)=0,
\end{equation}
where $x_i$, $i=0,1,2,3$, are the neighbours of $x$ in cyclic order and
\begin{equation}
v_{d_y}(x,x_i,x_{i+1}) := \min_{t\in[0,1]} t\,d_y(x_i) + (1-t)d_y(x_{i+1}) + \sqrt{\mu_{\delta}(x)} \sqrt{t^2 + (1-t)^2}.
\end{equation}
The solution to (\ref{eq:discreik}), which will be referred to as the \emph{eikonal distance}, together with the corresponding geodesics can be efficiently computed using a \emph{fast marching method} \cite{sethian_fast_1996,cohen_global_1997}.

\begin{figure}[t]
\begin{center}
\includegraphics[width=.32\linewidth]{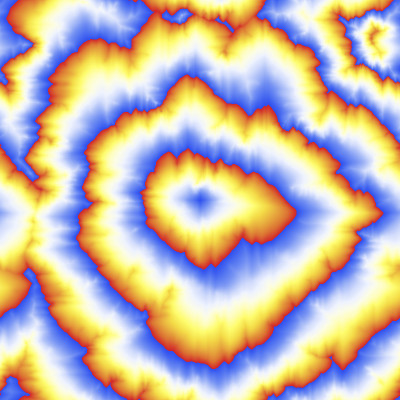}
\hspace{.1mm}
\includegraphics[width=.32\linewidth]{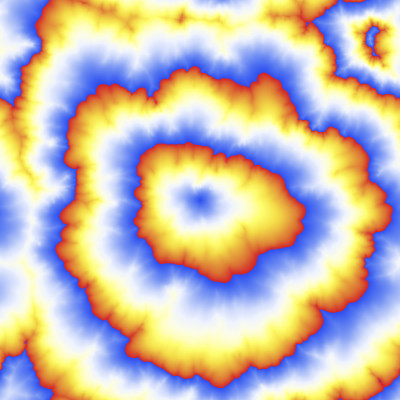}

\vspace{2.3mm}

\includegraphics[width=.32\linewidth]{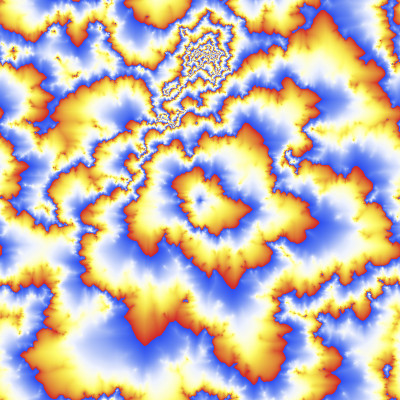}
\hspace{.1mm}
\includegraphics[width=.32\linewidth]{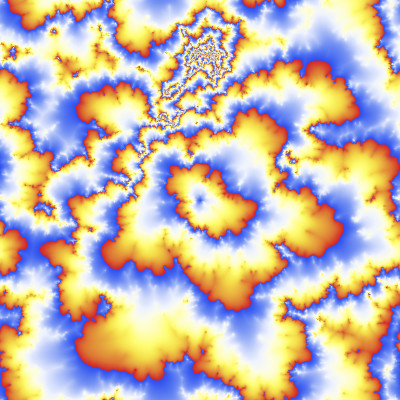}
\end{center}
\caption{The geodesic distance from the point at the center. For the left column the grid graph distance is used and the eikonal distance for the right column. Distances are computed w.r.t a disk-averaged quantum Liouville measure with $\gamma = 0.4$, $\delta=1/10000$ for the top row and $\gamma = 1$, $\delta=1/64000$ for the bottom row. }%
\label{fig:dist}
\end{figure}

In figure \ref{fig:dist} the distance $d_{\delta}(x,y)$ as a function of $x$ for a fixed point $y$ at the center of the lattice is plotted as a color gradient. 
The images on the left show the grid graph distance (\ref{eq:discrdist}) and on the right the eikonal distance.
The top two images correspond to a disk-averaged Liouville measure with $\gamma=0.4$ (and $\delta=1/10000$) showing an observable anisotropy for the grid graph distance, which is not present for the eikonal distance.
For larger $\gamma$ the effect is not so apparent, as can be seen in the bottom two images at $\gamma=1$, but still there is a systematic overestimation of the distances in the diagonal directions in the case of the grid-graph distance.

The solutions to the discrete Eikonal equation (\ref{eq:discreik}) for a measure approximating that of a Riemannian metric have been proven to converge to geodesic distances when the lattice size is so large that the relative difference between the measure of neighbouring lattice sites is much smaller than one (see e.g., \cite{rouy_viscosity_1992}).
In the case of a disk-averaged Liouville measure, the relative difference $(\mu_{\delta}(x)-\mu_{\delta}(x'))/\mu_{\delta}(x)$ for neighbouring lattice sites $x$ and $x'$ is by construction\footnote{Notice that the radius $\epsilon_x$ of $\epsilon_{x'}$ of neighbouring sites must differ by less than one, since otherwise one disk is a proper subset of the other and their measures cannot agree.} bounded by the inverse disk radius $1/\epsilon_{x}$.
We therefore trust the discrete geodesic distance to be a good approximation of (\ref{eq:dist}) in regions where $\epsilon_x$ is much larger than one. 
Not only do we not trust the geodesic distance formula in regions where $\epsilon_x$ is of order one, these are also the regions where one is probing the discrete quantum Liouville measure at the discretization scale, where it is known to deviate from the continuum quantum Liouville measure.

What does this mean in practice? 
Preferably one would choose $\delta$ and the lattice size $w$ such that the radius $\epsilon_x$ is uniformly bounded from below by some number larger than one. 
In particular, this would require that the measure at each lattice site is smaller than $\delta$, which is much too restrictive in practice.
Indeed, the maximum of a discrete Gaussian free field $h$ is known to grow with $h$ almost surely as $2 \log(w)$ (see \cite{bolthausen_entropic_2001}), which implies the bound $w^{-(\gamma-2)^2/2} \lesssim \delta$.
Only for very small values of $\gamma$, say $\gamma \lesssim 1/3$, can one satisfy this bound in practice while maintaining a reasonable range of allowed $\delta$'s.
In general, however, one will have to deal with the fact that the discrete geodesic distance is unreliable in regions where the measure per site is of the order $\delta$ (or greater).

\begin{figure}[t]
\includegraphics[width=.91\linewidth]{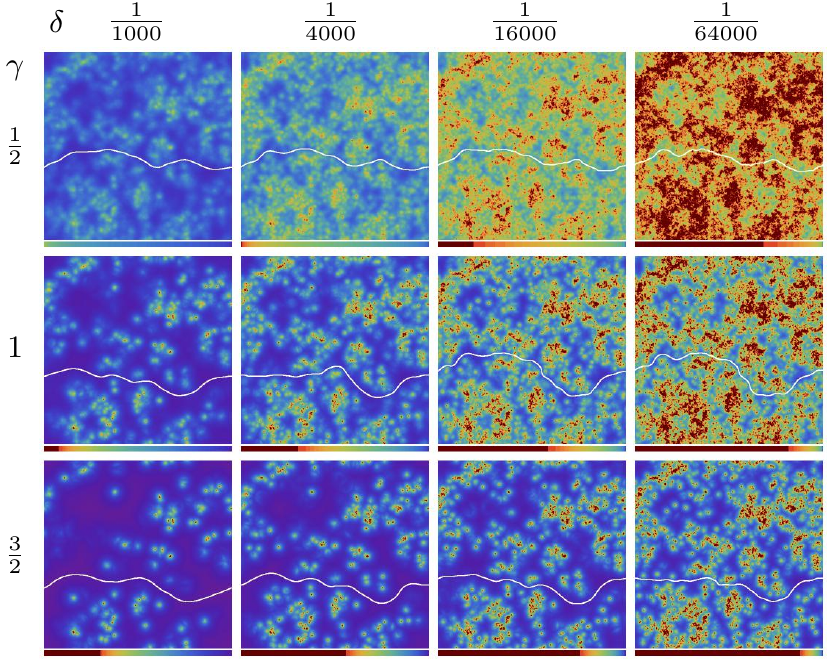}
\includegraphics[width=.08\linewidth]{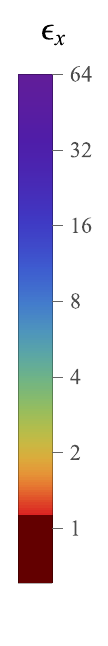}
\caption{The disk-averaged measures $\mu_{\gamma,\delta}$ corresponding to a single Gaussian free field on a lattice of size $w=512$ for various values of $\gamma$ and $\delta$. The legend on the right indicates the radius of averaging corresponding to the colors, while the bars beneath the images show how the measure is distributed over the lattice sites of varying color. The white curves indicate the shortest cycles discussed in section \ref{sec:cycles}.}%
\label{fig:cycles}
\end{figure}

In figure \ref{fig:cycles} a set of disk-averaged measures is shown corresponding to a single discrete Gaussian free field on a lattice of size $w=512$ for various values of $\gamma$ and $\delta$.   
The legend at the right indicates to what radius $\epsilon_x$ the colors correspond, showing in particular that the dark red lattice sites have measure $\mu_{\delta,\gamma}(x) > \delta/4$.
For most of the shown values of $\gamma$ and $\delta$ only a tiny fraction of the lattice sites is colored dark red.
However, typically those lattice sites together have non-negligible measure.
For instance, one can deduce from the colored bars beneath the images, which indicate how the measure is distributed among the various colors, that for $\gamma=1.5$ more than half of the measure is supported on the dark red spots for $\delta \lesssim 1/4000$. 

The skewed distribution of the measure especially for large $\gamma$ does not come as a surprise.
In fact, what one is seeing is the appearance of so-called \emph{$\gamma$-thick points} of the quantum Liouville measure.
A $\gamma$-thick point of a Gaussian free field $h$ is a point $x$ for which the average $h_{\epsilon}(x)$ of $h$ on a circle of radius $\epsilon$ centered at $x$ satisfies \cite{duplantier_liouville_2011}
\begin{equation}
\liminf_{\epsilon\to 0} \frac{h_{\epsilon}(x)}{-\log(\epsilon)} = \gamma
\end{equation}
In \cite{duplantier_liouville_2011} it was proved that a random point chosen with respect to a quantum Liouville measure $\rmd\mu_{\gamma}$ is almost surely a $\gamma$-thick point of the corresponding Gaussian free field.
Moreover, according to \cite{hu_thick_2010}, the set of $\gamma$-thick points is a fractal subset of the plane having Hausdorff dimension almost surely equal to $2 - \gamma^2/2$  (see \cite{daviaud_extremes_2006} for a similar result for the discrete Gaussian free field). 
Therefore, if we want a typical $\gamma$-thick point in the disk-averaged measure to have a radius $\epsilon_x$ larger than one, we get the bound $w^{\gamma^2/2-2} \lesssim \delta$, which for practical purposes is still too strict for large $\gamma$.

When it comes to defining observables associated to geodesic distances, one needs to take these considerations into account.
In the case of random triangulations an often studied observable is the two-point function corresponding to the probability density function of the geodesic distance between two randomly sampled vertices \cite{ambjorn_scaling_1995,ambjorn_fractal_1995}.
In terms of the quantum Liouville metric one would define the two-point function
\begin{equation}
G_{\gamma,\delta,V}(r) := \frac{1}{V^2}\left\langle\int \rmd \mu_{\gamma,V}(x)\int \rmd\mu_{\gamma,V}(y) \,\delta\left( r - d_{\delta}(x,y)\right)\right\rangle,
\end{equation}
corresponding to distribution of the geodesic distance $d_{\delta}(x,y)$ for randomly sampled points $x$ and $y$.
Since these points are almost surely $\gamma$-thick points, there is little hope of measuring $d_{\delta}(x,y)$ accurately over a large range of $\delta$'s.  

Alternatively, one could measure distances between points sampled according to the Lebesgue measure or a measure obtained from a non-trivial power of the quantum Liouville measure density.
However, in the case of the torus there exists an even simpler and well-defined geodesic distance that has a natural analogue in random triangulations, namely the length of the shortest non-contractible closed geodesic.

\section{Shortest cycles}\label{sec:cycles}

For a Riemannian metric $\rho(x)\rmd x^2$ on $\mathbb{T}^2$ with 
$\tau=i$, one can define the length $L$ of the shortest 
cycle in terms of the distance function $d(x,y)$ on its universal cover by
\begin{equation}
L := \inf_{x\in\R^2,\, a\in \Z^2\setminus\{0\}}  d(x,x+a).
\end{equation} 
Similarly, one may use either the grid graph distance (\ref{eq:discrdist}) or the solution to the discrete Eikonal equation (\ref{eq:discreik}) on the universal cover of the $w\times w$ lattice to define a discrete shortest cycle
\begin{equation}
L_{\gamma,w,\delta} := \inf_{x\in\Z^2,\, a\in \Z^2\setminus\{0\}}  d_{\delta}(x,x+w\,a).
\end{equation} 
Notice that it is not necessary to compute the distance $d_{\delta}(x,y)$ for each base point $y$.
Indeed, it suffices to compute it for base points $y$ taken from one row and one column of the lattice, since a non-contractible cycle is guaranteed to traverse at least one of these points.
In figures \ref{fig:reg} and \ref{fig:cycles} the shortest cycles are indicated by a solid white curve.
As expected, the shortest cycles stay clear of the $\gamma$-thick points as much as possible, and therefore we expect its length to be much less sensitive to lattice artifacts than geodesic distances appearing in the two-point function.
Of course, to really test for the absence of lattice artifacts one should study the scaling behaviour of $L_{\gamma,w,\delta}$ as $w\to\infty$ and $\delta\to 0$.
  
\begin{figure}[t]
\begin{center}
\includegraphics[width=.7\linewidth]{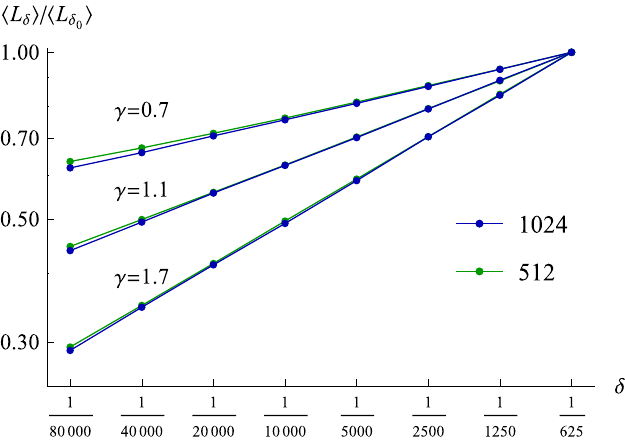}
\end{center}
\caption{The expectation values of the shortest cycle length $L_{\gamma,w,\delta}$ in terms of eikonal distance in a box-averaged quantum Liouville metric for various values of $\delta$, $\gamma$, and lattice sizes $w=512,1024$, normalized with respect to $\delta = \delta_0 = 1/625$. }%
\label{fig:lenscale}
\end{figure} 
 
\begin{figure}[t]
\begin{center}
\includegraphics[width=.49\linewidth]{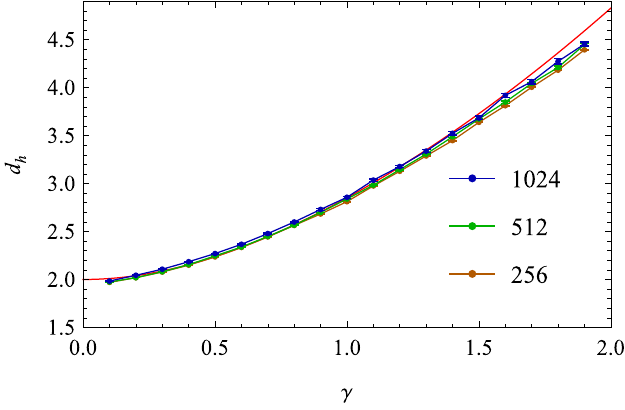}
\includegraphics[width=.49\linewidth]{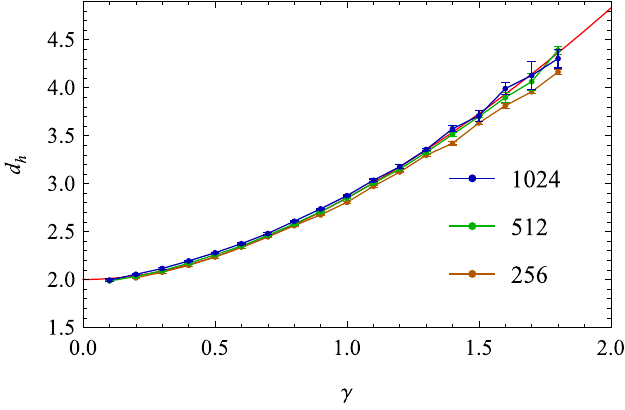}

\hspace{.06\linewidth}(a)\hspace{.47\linewidth}(b)
\end{center}

\vspace{-0.5cm}
\caption{The dimension $d_h$ as extracted from (a) the scaling of the expectation value of $L_{\gamma,w,\delta}$, and (b) the collapse of the probability density functions of $L_{\gamma,w,\delta}$. Formula (\ref{eq:watab}) is plotted in red. }%
\label{fig:hausd}
\end{figure}

In figure \ref{fig:lenscale} the expectation values $\langle L_{\gamma,w,\delta} \rangle$ are plotted for various values of $\gamma$, $\delta$ and $w$. 
These data points are for the eikonal distance in a box-averaged quantum Liouville metric without volume normalization, but other choices lead to very similar plots.
Only a very minor dependence on the lattice size is visible, while $\langle L_{\gamma,w,\delta} \rangle$ seems to scale as a power-law as function of $\delta$, in accordance with the conjecture (\ref{eq:distconv}).
By determining the power through a fit to the data in a suitable range of $\delta$'s, an estimate is obtained for the dimension $d_h$, which is shown in figure \ref{fig:hausd}a.
Figure \ref{fig:hausd}b shows the same dimensions $d_h$, but obtained instead by fitting, for fixed $w$ and $\gamma$, the probability density functions $P_{\gamma,\delta}(L)$ for a range of $\delta$'s to a (smooth interpolation of) a reference distribution $P_{\gamma,\delta_0}(L)$ for some fixed $\delta_0$,
\begin{equation}
P_{\gamma,\delta}(L) = (\delta/\delta_0)^{\frac{1}{d_h}-\frac{1}{2}} P_{\gamma,\delta_0}(L (\delta/\delta_0)^{\frac{1}{d_h}-\frac{1}{2}}).
\end{equation}
The data is seen to agree very well with formula (\ref{eq:watab}), which is shown in red in figure \ref{fig:hausd}. 

Similar scaling of lengths of shortest cycles in agreement 
with formula (\ref{eq:watab}) was previously observed for random 
triangulations, both in pure gravity \cite{ambjorn_baby_2011}  
and for gravity coupled to matter fields with central charge $c=-2$ \cite{ambjorn_roaming_2011}.\footnote{Also, similar scaling was observed for second-shortest closed geodesics in random triangulations coupled to an Ising model and a 3-state Potts model \cite{ambjorn_toroidal_2013}.} 
In the latter case random triangulations are sampled uniformly from the ensemble of triangulations of the torus with a marked spanning tree, which are objects that can be generated very efficiently (see \cite{ambjorn_roaming_2011} for details). 
This raises the question whether a more direct comparison might be possible between the distances in Liouville quantum gravity and random triangulations.
If both models live in the same universality class and our definition of distance is the appropriate one, the geodesic distances should agree up to an unphysical overall rescaling.
However, before comparing the data one should make sure that one is really comparing corresponding quantities.
Remember that we conditioned on modulus $\tau = i$ to simplify the discretization of the Gaussian free field, while generally there is no restriction on the (discrete) modulus of a random triangulation.
In principle, the numerical methods described in section \ref{sec:dgff} and \ref{sec:geod} can be straightforwardly generalized to $\tau \neq i$, but the result is fairly messy and prone to systematic discretization effects.
Instead, it is simpler to impose the condition $\tau \approx i$ on the random triangulations. 

\begin{figure}[t]
\begin{center}
\includegraphics[width=.56\linewidth]{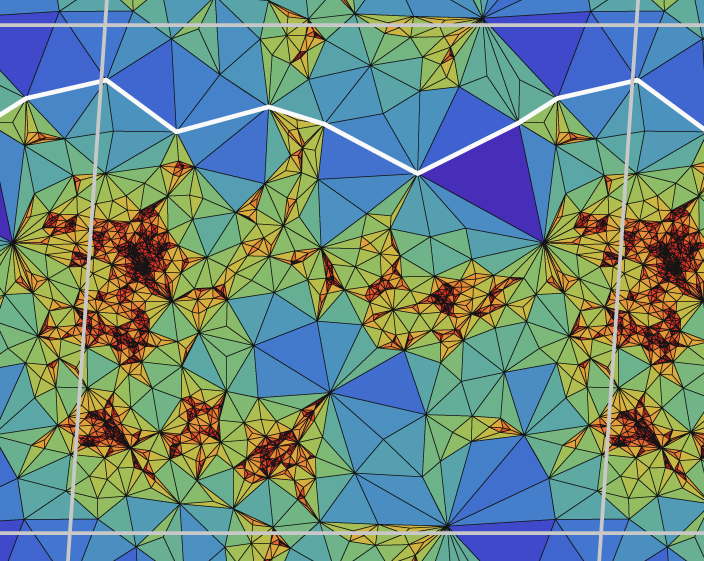}
\hspace{.05\linewidth}\includegraphics[width=.2\linewidth]{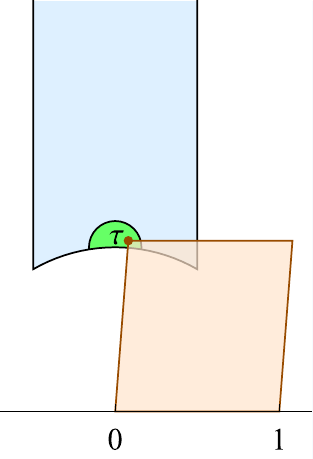}
\end{center}
\caption{ Example of a shortest loop in a random spanning-tree-decorated triangulation with $10000$ triangles conditioned to have discrete modulus $\tau$ satisfying $|\tau - i| < 0.16$ (see right figure). }%
\label{fig:dtloop}
\end{figure}

\begin{figure}[t]
\begin{center}
\includegraphics[width=.75\linewidth]{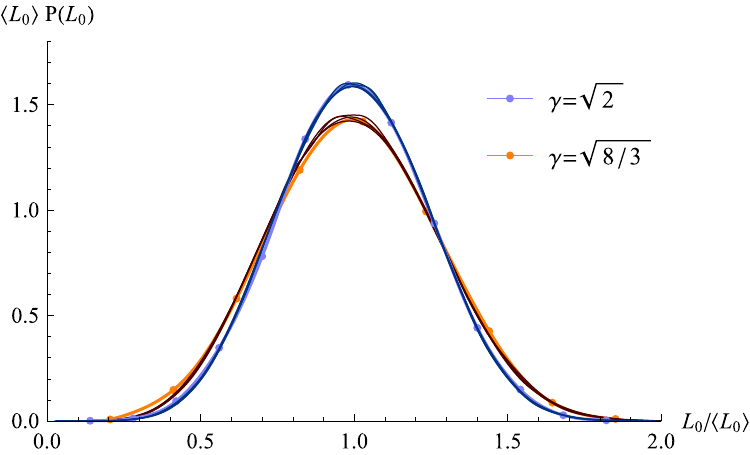}
\end{center}
\caption{Comparison to shortest cycles in random triangulations of the torus. The thick light curves represent the distributions of the shortest cycle in a uniform random triangulation of the torus (orange/broader) and a uniform random spanning-tree-decorated triangulation (blue/narrower) with $10000$ triangles.  The thin, darker curves represent the distributions of shortest cycle lengths in terms of the Gaussian free fields on a $256\times 256$-lattice. }%
\label{fig:compare}
\end{figure}

We have generated a large number of random triangulations, both undecorated and spanning-tree-decorated, with $10000$ triangles.
For each triangulation the discrete modulus $\tau$ was computed and all triangulations were discarded for which $\tau$ was too far from $i$ ($|\tau - i|<0.16$, to be precise, see figure \ref{fig:dtloop}).
The probability density functions $P(L)$ of the shortest cycles in these triangulations relative to their expectation values are plotted (in light-blue and orange) in figure \ref{fig:compare}. 

These are compared to the distributions $P_{\gamma,\delta}(L)$ of the shortest cycles in disk-averaged quantum Liouville measures (including the volume-normalization (\ref{eq:discrvolnorm})) with $\gamma = \sqrt{2},\sqrt{8/3}$. 
The data sets for $w=256$ and $\delta$ ranging from $1/80000$ to $1/10000$ are represented in figure \ref{fig:compare} by the thin, dark curves.
Observe that especially the data for $\gamma=\sqrt{2}$ agrees very well with that of the spanning-tree-decorated triangulations.
At the very least, one may conclude that in both cases the relative standard deviation of the length of the shortest cycles in the quantum Liouville measures accurately match those for the random triangulations.  

\section{Conclusions}

Physicists often speak of the ``quantum geometry'' of space-time which must arise from quantizing the dynamical metric in general relativity, but typically a good mental picture of such a geometry is lacking.
Two-dimension Euclidean gravity (coupled to matter 
with central charge $c\leq 1$) provides an explicit realization of quantum geometry that can be studied in detail. 
The fact that one can make sense of quantum geometry in two dimensions is not because it is any less quantum than it is in higher dimensions.
In fact, in some sense 
2d quantum gravity is maximally quantum, when defined in terms of the Einstein-Hilbert
action on a surface of fixed topology. 
Since the Einstein-Hilbert action is topological in 2d, fixing the topology means that there is 
no real action: in the path integral all geometries carry equal weight, which corresponds formally to the limit $\hbar \to \infty$. 
In such a setting of wildly fluctuating geometries it seems a priori not guaranteed at all that a notion exists that qualifies as ``quantum geodesic distance''. 
However, it was shown in the setting of random triangulations, first in \cite{kawai_transfer_1993} and extended in many directions both in the physics \cite{ambjorn_scaling_1995,ambjorn_fractal_1995} and mathematical literature \cite{chassaing_random_2004,marckert_limit_2006,bouttier_geodesic_2003,gall_uniqueness_2013,miermont_brownian_2013}, that a well-defined geodesic distance exists in the continuum.
Around the same time first steps were taken to identify a similar notion in Liouville quantum gravity 
\cite{david_what_1992,watabiki_analytic_1993}, but progress on this front has been limited until recently. 
The rigorous definition of the quantum Liouville measure \cite{duplantier_liouville_2011} now provides a firm basis and there is good hope that studying various processes coupled to this measure, most notably Brownian motion \cite{garban_liouville_2013,maillard_liouville_2014,rhodes_heat_2014} and quantum Loewner evolution \cite{miller_quantum_2013}, will shed light on its geometry.

In the setting of random triangulations, numerical simulations have on many occasions in the past decades served as a guiding principle and have lead to many new conjectures, some of which have later been proven rigorously.
In this paper we have attempted to extend the numerical toolbox to Liouville quantum gravity, where hopefully it will serve a similar purpose.
The presented results indicate that, as far as the lengths of shortest cycles are concerned, the proposed definition of geodesic distance seems to provide a well-defined metric structure.
Moreover, the simulations provide numerical evidence in favour of the conjectured formula (\ref{eq:watab}) for the Hausdorff dimension, and quantitative agreement with distances in random triangulations is found.
We believe that the numerical tools can be applied more widely than just to the shortest cycles considered in this paper, but, as detailed in section \ref{sec:geod}, due to the fractal nature of the measure great care is required to avoid lattice artifacts in the data.

\subsection*{Acknowledgments}

The authors acknowledge support from the ERC-Advanced grant 291092,
``Exploring the Quantum Universe'' (EQU). JA acknowledges support
of FNU, the Free Danish Research Council, from the grant
``quantum gravity and the role of black holes''.   
 In addition JA was supported 
in part by Perimeter Institute of Theoretical Physics.
Research at Perimeter Institute is supported by the Government of Canada
through Industry Canada and by the Province of Ontario through the 
Ministry of Economic Development \& Innovation.
 
\appendix
\section{Source code}

The simulation software used to obtain the reported data was written in C++. 
To generate discrete Gaussian free fields a Fast Fourier Transform library, called FFTW \cite{frigo_fast_1999}, was used.
The source code is available at: 

\url{https://github.com/tgbudd/Lattice-Liouville}

\bibliographystyle{habbrv}
\bibliography{liouville}

\end{document}